\documentclass[twocolumn,aps,prb,superscriptaddress,showpacs,amsmath,amssymb,floats]{revtex4}

\usepackage{amsmath}
\usepackage{amssymb}
\usepackage{graphicx}
\usepackage{keyval}
\usepackage{dcolumn}
\usepackage{bm}
\usepackage{color}
\usepackage{txfonts}
\usepackage[]{sidecap}

\begin{document}

\title{The origin of the dead-layer at the La$_{0.67}$Sr$_{0.33}$MnO$_3$/SrTiO$_3$ interface and dead-layer reduction via interfacial engineering}

\author{R. Peng}\author{H. C. Xu}\author{M. Xia}\author{J. F. Zhao}\author{X. Xie}\author{D. F. Xu}\author{B. P. Xie}\author{D. L. Feng}\email{dlfeng@fudan.edu.cn}
\affiliation{State Key Laboratory of Surface Physics, Department of
Physics, and Advanced Materials Laboratory, Fudan University,
Shanghai 200433, People's Republic of China}

\date{\today}

\begin{abstract}
\textbf{Transition metal oxide hetero-structure has great potential for multifunctional devices. However, the degraded physical properties at interface, known as dead-layer behavior, present a main obstacle for device applications. Here we present the systematic study of the dead-layer behavior in La$_{0.67}$Sr$_{0.33}$MnO$_3$ thin film grown on SrTiO$_3$ substrate with ozone assisted molecular beam epitaxy. We found that the evolution of electric and magnetic properties as a function of thickness shows a remarkable resemblance to the phase diagram as a function of doping for bulk materials, providing compelling evidences of the hole depletion in near interface layers that causes dead-layer. Detailed electronic and surface structure studies indicate that the hole depletion is due to the intrinsic oxygen vacancy formation. Furthermore, we show that oxygen vacancies are partly caused by interfacial electric dipolar field, and thus by doping-engineering at the single-atomic-layer level, we demonstrate the dead-layer reduction in films with higher interfacial hole concentration.}
\end{abstract}
\pacs{75.47.Lx, 75.30.Kz, 79.60.Dp}

\maketitle

The complexity in transition metal oxides and their hetero-structures, due to the entangled correlation of charge, spin, orbital, and lattice degrees of freedom, is a double-edged sword. On one hand, it brings out emergent phenomena in condensed matter physics \cite{LTOSTO,LAOSTO,LAOSTO2,LAOSTO3} and various possibilities for multifunctional device applications \cite{app1,app2}. On the other hand, it often conceals the physical mechanism of new phenomena due to theoretical difficulties and hinders the solutions to problems in device applications. One typical example is La$_{0.67}$Sr$_{0.33}$MnO$_3$/SrTiO$_3$ (LSMO/STO) hetero-structure. LSMO is well-known for the colossal magnetoresistance, half metallic behavior, room temperature ferromagnetism, and high conductivity. These exotic properties make LSMO the most promising material for metal-based spintronic device, magneto-tunneling junctions, magnetic memory, \textit{etc}. Success has been achieved in developing LSMO-based field effect transistors \cite{LSMOapp2} and metal-base transistors \cite{LSMOapp3}. However, in spite of a few reported working devices, a large variety of applications are restricted due to the dead-layer behavior, that is, the degraded ferromagnetism and metallicity with decreasing thickness and eventually insulating below certain critical thickness \cite{DLdiagram,deadlayer5}. Therefore, it is crucial to investigate the mechanism of the dead-layer behavior in LSMO/STO, for both device application and fundamental understanding of oxide interface.

\begin{figure}[t]
 \includegraphics[width=8cm]{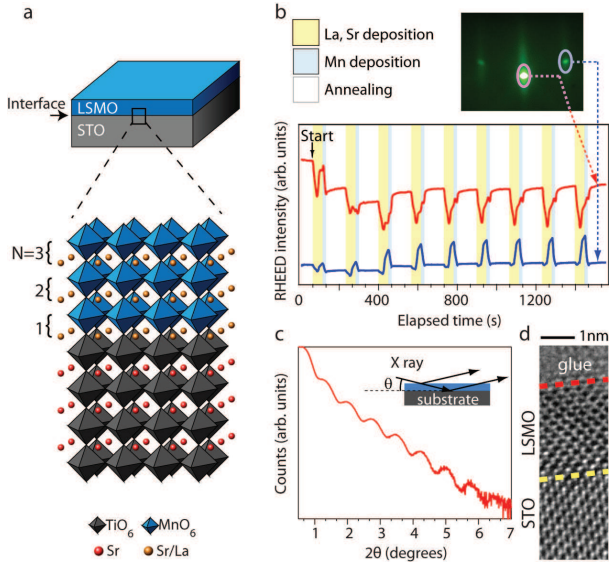}
\caption{\textbf{Atomic layer-by-layer growth of LSMO/STO.}  \textbf{a}, Schematic illustration of the films studied, and $N$ numbers the different layers around interface. \textbf{b}, The RHEED intensity oscillations of the specular spot (red line) and diffracted spot (blue line) during the deposition process shown by the colored background. The RHEED pattern is retained two dimensional after this 7~u.c. LSMO film growth as shown. \textbf{c}, X-ray reflection measurement on a 30~u.c. LSMO film in order to calibrate the thickness of the films grown successively within one comparative set. The geometry of the X-ray reflection is shown in the inset. By fitting the modified Bragg equation \cite{XRR}, the actual thickness of this film was calculated to be 29.8~u.c. \textbf{d}, Cross-sectional TEM image of 7~u.c. LSMO/STO off-normal direction. The yellow and red dashed lines indicate the interface and thin film surface, respectively.}
\label{intro}
\end{figure}

Intensive studies have been conducted, and several scenarios have been proposed to explain the dead-layer behavior, such as magnetic \cite{deadlayer1,Mnvalence} and orbital reconstruction at the interface \cite{deadlayer2,deadlayer3,deadlayer4}, and substrate-induced strain \cite{strain,deadlayer5}. However, the situation is still very perplexing, and many issues remain to be understood. For example, it is not clear why the complicated electric and magnetic properties in ultra-thin films are extremely sensitive to film thickness. Extrinsic imperfections induced during fabrication complicate the problem even further \cite{deadlayer6}. Here we study the dead-layer behavior of LSMO/STO by ozone-assisted molecular beam epitaxy (OMBE), where the low-energy growth mode reduces extrinsic imperfections to the lowest level, and precise control of the composition becomes possible at the single-atomic-layer level \cite{OMBE1,OMBE2}.

\begin{figure}[b]
\includegraphics[width=8cm]{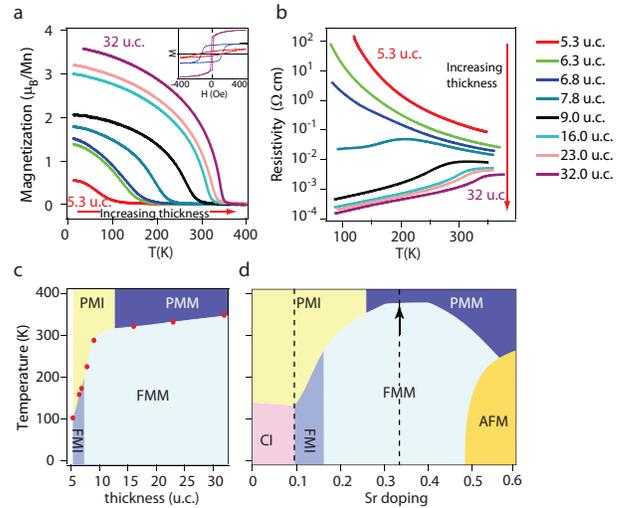}
\caption{\textbf{The physical properties of the La$_{0.67}$Sr$_{0.33}$MnO$_3$ films with various thicknesses.} \textbf{a}, Temperature dependence of the magnetization for films with various thicknesses measured at 1000~Oe along the (100) direction. Samples are field-cooled from 400~K. The inset shows the hysteresis loops measured at 10~K for 5.3~u.c., 6.8~u.c. and 32~u.c. films, indicating the ferromagnetism of all these samples. \textbf{b}, Temperature dependence of resistivity for films with various thicknesses. \textbf{c}, Phase diagram of LSMO/STO as a function of film thickness according to the data in \textbf{a} and \textbf{b}. \textbf{d}, Reproduced phase diagram of La$_{1-x}$Sr$_x$MnO$_3$ bulk material as a function of Sr substitution for La (ref. \onlinecite{phasediagram}). The abbreviations stand for spin-canted insulator (CI), ferromagnetic insulator (FMI), paramagnetic insulator (PMI), ferromagnetic metal (FMM), paramagnetic metal (PMM) and antiferromagnetic metal (AFM), respectively. The temperature scale is the same with \textbf{c}. The arrow shows the doping of the LSMO we studied, and the region between the dashed lines shows a remarkable resemblance to \textbf{c}.}
\label{thicknessdep}
\end{figure}

Atomic layer-by-layer growth of LSMO was performed on TiO$_2$ terminated STO~(001) substrate as shown in Fig.~1. After optimizing film stoichiometry and growth condition, stable reflection high-energy electron diffraction~(RHEED) oscillations and two dimensional RHEED pattern are retained during growth (Fig.~1b). Rutherford backscattering spectrometry measurements were performed to further calibrate the stoichiometry. Films within one comparative set were fabricated successively in several hours under the same condition, together with a 30 unit cell (u.c.) film for sample quality verification and thickness calibration. X-ray reflection measurements were performed on the 30~u.c. LSMO films (Fig.~1c) to calibrate the thickness of thin films in the same comparative set, which is crucial for the critical thickness study. The cross-sectional transmission electron microscope~(TEM) image of the 7~u.c. LSMO (Fig.~1d) also confirms the accurate thickness. With precise stoichiometry and thickness control, films with various thicknesses and engineered carrier doping at individual atomic layers, were fabricated and investigated.

In Figs.~2a and 2b, the 32~u.c. LSMO shows a Curie transition at 350~K, close to the bulk value~(370K). With decreasing film thickness, the Curie temperature ($T_c$) drops, and the metallicity is suppressed. Below 7~u.c., the critical thickness, the film is insulating below $T_c$, demonstrating a typical dead-layer behavior. Figure~2c summarizes various phases in the thin films as a function of thickness according to data in Figs.~2a and 2b. Surprisingly, this phase diagram shows a remarkable resemblance to that of bulk La$_{1-x}$Sr$_{x}$MnO$_3$ crystals with varying Sr concentration \cite{phasediagram,phasediagram2}, as reproduced in Fig.~2d. With no Sr doping, LaMnO$_3$ is an antiferromagnetic insulator due to superexchange interactions. By introducing holes with Sr doping, the double-exchange interaction between the neighboring $3d^{3}$ and $3d^{4}$ Mn sites favors the spin-aligned ferromagnetic state. In lightly doped La$_{1-x}$Sr$_{x}$MnO$_3$, the sparse and inhomogeneous hole distribution results in small ferromagnetic metallic clusters embedded in insulating regions, while the system remains as an insulator with weak ferromagnetism and degraded conductivity below $T_c$ (ref. \onlinecite{phasediagram}). With sufficient holes, the ferromagnetic clusters eventually overlap, and thus it becomes a metallic ferromagnet through a percolation transition \cite{phasediagram}. The intriguing similarity in the two phase diagrams presented in Figs.~2c and 2d suggests that they probably have the same origin. It suggests that the hole concentration in an ultra-thin film decreases with decreasing thickness, which induces the insulating behavior below a critical thickness through the percolation transition.

\begin{figure}[t]
\includegraphics[width=8cm]{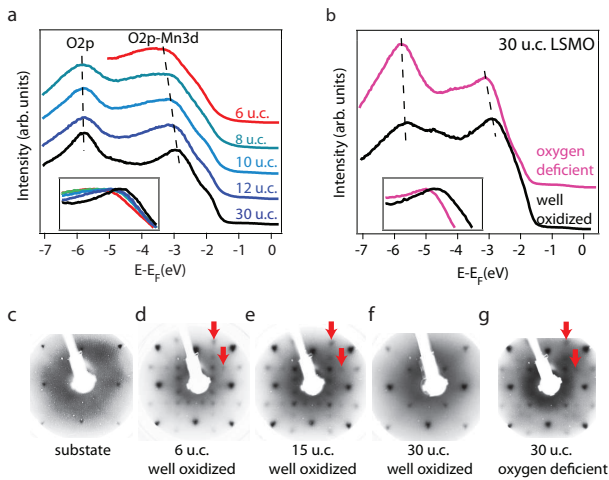}
\caption{\textbf{\textit{In situ} photoemission and LEED measurements on films with different thicknesses and growth environment.} \textbf{a}, Thickness dependence of \textit{in situ} angle-integrated photoemission spectra (AIPES) for LSMO/STO around the Brillouin zone center. The integrated EDC's are vertically offset. All data were taken with 21.2~eV photon energy at 25~K except for 6~u.c. film which is measured at 180~K to avoid charging effect. The prominent feature of valence band shifts towards higher binding energy with decreasing thickness, and the same evolution happens around the Brillouin zone edge. \textbf{b}, AIPES of 30~u.c. well oxidized LSMO/STO and the 30~u.c. LSMO/STO grown at oxygen deficient environment (under $5\times 10^{-7}$~mbar O$_2$ pressure) around the Brillouin zone edge. The inset in \textbf{a} and \textbf{b} show the O$2p$-Mn$3d$ hybridized band normalized by peak height. The valence band evolution due to oxygen deficiency is similar to that with decreasing film thickness. \textbf{c}-\textbf{g}, \textit{In situ} LEED pattern for STO substrate, well-oxidized 6~u.c. LSMO, well-oxidized 15~u.c. LSMO, well oxidized 30~u.c. LSMO and intentionally-oxygen-deficient 30~u.c. LSMO, respectively. The red arrows mark the additional reconstruction spots in the LEED patterns.} \label{insitu}
\end{figure}

If the hole depletion in the LSMO thin films had been caused by reduced Sr concentration near the interface, there should have been more than 0.15 Sr doping reduction for the 7~u.c. thick film by comparing the two phase diagrams. Moreover, Sr concentration would have changed continuously over a wide range of thickness to reproduce the transport properties. However, the RHEED oscillations became very stable just after the growth of the first 3~u.c. LSMO (Fig.~1b). Because the RHEED oscillations are sensitive to less than 1\% stoichiometry difference, it ensures the stoichiometry of La and Sr. More evidences come from the previous scanning transmission electron microscopy (STEM) studies on LSMO/STO superlattice that have never resolved such large cation non-stoichiometry \cite{deadlayer6}. Based on these arguments, we conclude that the reduction of the hole concentration cannot come from the La/Sr ratio variation.

To understand the resemblance between thickness dependence in ultra-thin LSMO and doping dependence in bulk La$_{1-x}$Sr$_{x}$MnO$_3$, we performed \textit{in situ} photoemission measurements (h$\nu$=21.2eV) on the LSMO films around critical thickness. In Fig.~3a, the prominent features of valence bands, assigned as O$2p$ band and O$2p$-Mn$3d$ hybridized band \cite{VB1}, shift systematically towards higher binding energy with decreasing thickness. Similar shift is also observed in the 30~u.c. film intentionally grown under oxygen deficient condition (Fig.~3b). This valence band evolution due to oxygen deficiency in LSMO is further supported by recent hard X-ray measurement reported on LSMO films annealed in vacuum \cite{VB2}. Therefore, our data indicate that below 30~u.c., oxygen vacancies increases with decreasing thickness. Additional support comes from the \textit{in situ} low-energy electron diffraction~(LEED) pattern (Figs.~3c-3g). The well-oxidized 30~u.c. film shows only ($\sqrt{2} \times \sqrt{2}$)R45$^{\circ}$ reconstruction compared with the substrate, known as the tilt and distortion of oxygen octahedral in manganite \cite{LEED}. While both the ultra-thin films and the intentionally-oxygen-deficient 30~u.c. film show an additional surface reconstruction (the red arrows in Figs.~3c-3g). We emphasize that all films were grown in the oxygen adequate atmosphere if not specified otherwise, and further annealing in ozone did not cause any variation in the valence band or LEED patterns. Therefore, oxygen vacancies are somehow favored energetically in the ultra-thin films, which break the double exchange bonding between Mn$^{3+}$ and Mn$^{4+}$ and induce electron doping. As a result, the ferromagnetism can be severely degraded even with a few oxygen vacancies, which are likely not resolved in the RHEED oscillation. Therefore, based on the photoemission data and LEED pattern, we conclude that oxygen vacancies are the major source for the extra electrons transferred to Mn, which decrease the hole concentration. It thus well explains the similarity of the two phase diagrams shown in Fig.~2.

\begin{figure*}[t]
 \includegraphics[width=17cm]{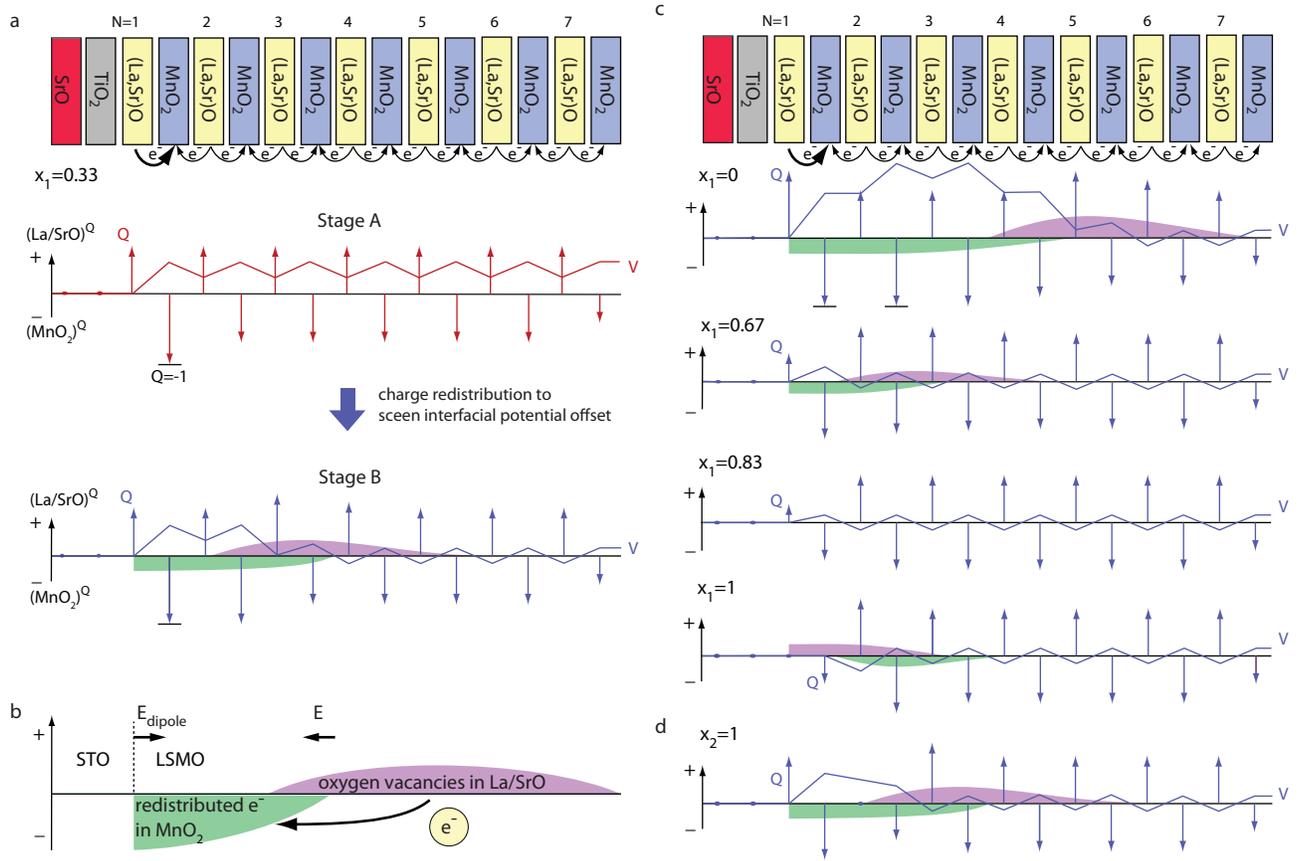}
\caption{\textbf{Schematic illustration of charge redistribution and intrinsic oxygen vacancy formation in LSMO/STO.} \textbf{a}, The electric potential and charge distribution in LSMO/STO film. Polar catastrophe is naturally avoided by varying the net charge in MnO$_2$ layers at the interface and surface layers (red arrows) in stage A, however, an electric potential offset remains between substrate and thin film (red curve). In order to screen this potential offset, the charges redistributes to stage B (blue arrows), which prefers oxygen vacancies in La/SrO layer (purple area) and decreased hole doping in MnO$_2$ layer near interface (green area). After this redistribution, the electric potential in thin film approaches the potential of the substrate within several layers (blue curve). \textbf{b}, The schematic diagram of the oxygen vacancies and charge redistribution caused by the electric potential offset at interface. The induced electric field direction is opposite to the interface dipole, thus compensating the interfacial potential difference. \textbf{c}, Evolution of oxygen vacancies and charge redistribution with interfacial doping $x_1$=0, 0.67, 0.83, 1, respectively, which tune the interfacial potential offset. \textbf{d}, The oxygen vacancies and charge redistribution in the case of $x_1$=0.33 and $x_2$=1.}
\label{model}
\end{figure*}

Since the oxygen vacancies are ordered as observed by LEED, it is likely due to certain energetically favored intrinsic process. To fully understand this, rather sophisticated \textit{ab initio} calculations are needed for such correlated materials. However, one could still explore with simple electrostatic model to study how the interface causes spontaneous charge redistribution in thin films. A well-known example is the LaAlO$_3$ film on STO substrate~(LAO/STO), where the intriguing high mobility carriers at the interface were well explained by a simple electrostatic model of the charge redistribution driven by electrostatic potential \cite{polarcata}. With a similar electrostatic model shown in Fig.~4, we schematically study the charge redistribution in LSMO/STO interface. The polar catastrophe is naturally avoided as shown by stage A in Fig.~4a, which is identical to the LAO/STO case except at the interface. Since the lowest unoccupied state in STO (Ti $t_{2g}$) is about 1eV higher than that of the thin film (Mn $e_g$) \cite{robustTi}, the (La,Sr)O layer at the interface tends to give all its charges (0.67e$^{-}$ on average) to the neighboring MnO$_2$ layer other than the TiO$_2$ layer, leaving a potential offset between substrate and film. To maintain the same electric potential between the substrate and surface, internal charge redistribution in LSMO film has to take place. With fixed La/Sr stoichiometry, a likely reaction of the system is to create oxygen vacancies in the (La,Sr)O layers, which effectively allows more electrons to be transferred to the MnO$_2$ layers near interface, and makes the (La,Sr)O$_{1-\delta}$ layers more positively charged as illustrated by Stage B of Fig.~4a. Figure~4b schematically illustrates the potential-induced electron redistribution. The induced electric field direction is opposite to the interface dipolar field, thus compensating the interfacial electric potential. As a result, Mn$^{3+}$ characters are enhanced near the interface, which was indeed found before in the X-ray absorption study on LSMO grown by pulsed laser deposition \cite{Mnvalence}.

If our model does capture certain essence of what happens near the interface, varying the interface dipole by interface engineering should substantially affect the dead-layer behavior. Several designs are laid out in Fig.~4c. For the film with $x_1$=0 ($x_N$ stands for the engineered Sr doping $x$ at the $N$'th layer), the interface induces larger electric potential difference, and the compensating process requires more redistributed electrons and oxygen deficiencies over a larger spatial range. For the film with $x_1$=0.67, the electric potential is reduced, therefore, fewer oxygen vacancies will be induced. While for the film with $x_1$=0.83, there is no electrostatic potential offset, and thus no induced oxygen vacancy under the electrostatic consideration. For the film with $x_1$=1, the interface dipole is opposite with the unengineered film shown in Fig.~4a, resulting in the confined oxygen vacancies near the interface. Since the oxygen vacancies are less widespreading in the films with higher $x_1$, these films should have enhanced ferromagnetism and metallicity according to our scenario, while the film with $x_1$=0.83 should have the most reduced dead-layer. Moreover, it does not help to balance the interfacial dipole if the additional Sr is doped away from interface, for example, in the $x_2$=1 case shown in Fig.~4d, therefore the dead-layer behavior should not be effectively suppressed.

\begin{figure*}[t]
\includegraphics[width=14cm]{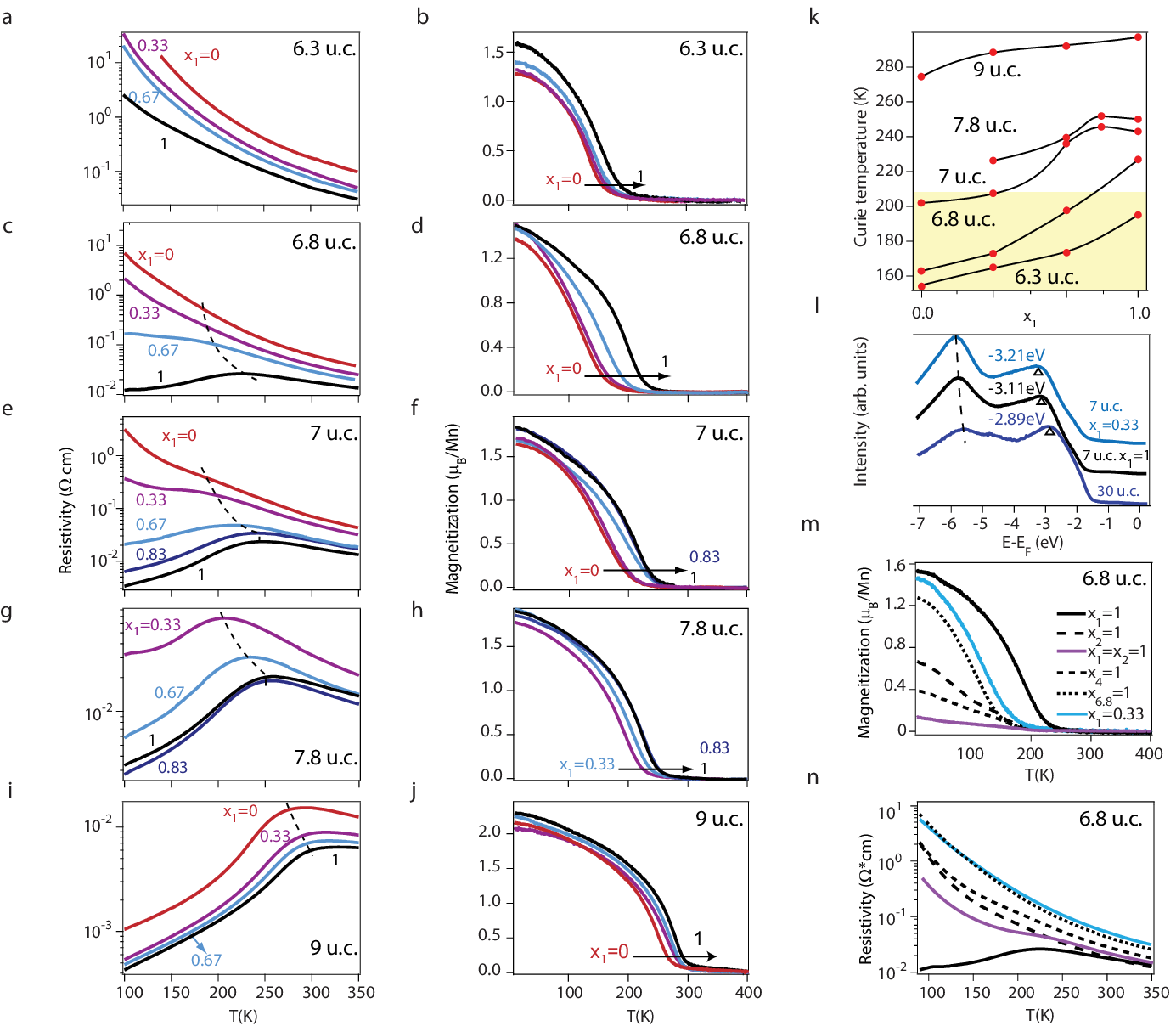}
\caption{\textbf{Interface engineering on LSMO/STO.} \textbf{a}, Resistivity as a function of temperature for 6.3~u.c. LSMO films with various $x_1$. \textbf{b}, Magnetization as a function of temperature for 6.3~u.c. LSMO films with various $x_1$. \textbf{c}, \textbf{e}, \textbf{g}, \textbf{i}, same as \textbf{a}, but for 6.8~u.c., 7~u.c., 7.8~u.c., 9~u.c. LSMO films, respectively. \textbf{d}, \textbf{f}, \textbf{h}, \textbf{j}, same as \textbf{b}, but for 6.8~u.c., 7~u.c., 7.8~u.c., 9~u.c. LSMO films, respectively. \textbf{k}, Summary of the Curie temperatures for films with various $x_1$ and thickness. The films in the yellow area are insulating in the entire temperature range. \textbf{l}, AIPES around the Brillouin zone edge for 7~u.c. LSMO films with different $x_1$, and a 30~u.c. unengineered film. Films are grounded by graphite on the edge to avoid the energy shift due to Schottky barrier reported in La$_{1-x}$Sr$_x$MnO$_3$/Nb:SrTiO$_3$ with SrO interface \cite{barrier}. \textbf{m}, Resistivity and \textbf{n}, magnetization of the films with $x_N$=1 at the $N$'th layer. Data of the unengineered $x_1$=0.33 film is also plotted for comparison. High Sr doping is only effective at the interfacial layer for the purpose of dead layer reduction.}\label{engineering}
\end{figure*}

With OMBE, one can effectively test the above predictions. We fabricated films with engineered $x_1$. As shown in Figs.~5a-5f, the resistivity decreases and Curie temperature increases with higher $x_1$ for 6.3~u.c., 6.8~u.c. and 7~u.c. thick films, which are below the critical thickness of the unengineered films. Curie temperatures are enhanced by 30~K, 54~K, 38.5~K compared with the unengineered films for 6.3~u.c., 6.8~u.c. and 7~u.c. thickness, respectively. Especially, the $x_1$=0.83 and $x_1$=1 films are metallic for 6.8~u.c. and 7~u.c., indicating that the dead-layer is successfully reduced to 6~u.c. While for the 7.8~u.c. and 9~u.c thick films (just above critical thickness), the Curie temperatures are also enhanced by 25.5~K and 9~K respectively, when comparing the unengineered ($x_1=0.33$) and engineered ($x_1$=1) films; their low temperature resistivities are reduced to 7.8\% and 78.8\% of their unengineered values, respectively (Figs.~5g-5j). Fig.~5k summarizes the Curie temperatures as a function of $x_1$ and film thickness. It indicates that the deteriorated metallicity and ferromagnetism with decreased thickness can be compensated by higher interfacial Sr doping. The Curie temperatures of all the thin films reach their maxima when $x_1$=0.83 or 1, as predicted in Fig.~4c. The shift of valence band towards lower binding energy with higher $x_1$ (Fig.~5l) further confirms that the interface doping reduce oxygen vacancies throughout the film.

It is important to note that simply adding more holes by increasing Sr concentration does not suppress the dead-layer behavior. As predicted in Fig.~4d and experimentally illustrated in Figs.~5m and 5n, higher Sr doping is only effective at the interfacial layer in reducing the dead-layer behavior. While doping at other atomic layer, or make $x_1$=$x_2$=1 could neither improve the Curie temperature nor bring out metallic behavior at low temperature in the 6.8~u.c. films, confirming the interfacial potential difference as a driving force for oxygen vacancy formation, which subsequently causes dead layer.

The predicting power of our simple model indicates that it does capture some essence of the dead-layer phenomenon. On the other hand, we note that it still could not explain why dead-layer still exist in the electrostatic balanced $x_1$=0.83 film. As shown in Fig.~5l, the O$2p$-Mn$3d$ hybridized band position of the engineered film shifts 0.10~eV towards that measured on the 30~u.c. film, but there is still a difference of 0.22~eV. That is, judging from the band shift, there are still about 70\% oxygen vacancies left in the engineered film. As a result, the interface-engineered 6.3~u.c. films are still insulating. Moreover, considering the valence band shift in 12~u.c. film (Fig.~3a) and additional superstructure of 15~u.c. film (Fig.~3e), the oxygen vacancies should be more spatially widespreading than illustrated. Furthermore, in oxygen-deficient epitaxial LSMO films, the previous x-ray photoemission studies have found evidence for oxygen vacancies in the MnO$_2$ layers \cite{VB1}. Therefore, Mn valence is decreased over a large spacial range, not only by inter-layer electron transfer due to the oxygen vacancies in the La/SrO layer, but also by the oxygen vacancies in the same MnO$_2$ layer. For the MnO$_6$ octahedral in a simple cubic structure, the oxygen vacancies in the La/SrO layer and the MnO$_2$ layer are equivalent, and both cases could conspire dead-layer behavior. However, the latter case leaves the net charge unchanged, and thus might not be directly induced by the interfacial dipole. These suggest that although we have significantly reduced the oxygen vacancies with the engineered interface, there are other factors for the oxygen vacancy formation that is beyond our simple model. More comprehensive \textit{ab initio} calculations are needed to better illustrate why oxygen vacancies are energetically favored. Dead-layer could hopefully be further suppressed with atomic engineering if oxygen vacancy formation is further understood.

To summarize, we have shown that dead-layer formation at LSMO/STO interface is due to the hole depletion by intrinsic oxygen vacancies. It gives a full account for the electric and magnetic properties and complex phase diagram of the ultra-thin films. Moreover, we proposed a dead-layer scenario, which suggests the charge redistribution and intrinsic oxygen vacancies are partly induced by interface potential difference. We have predicted and confirmed experimentally that the dead-layer behavior could be weakened through atomic scale engineering of the interface, and achieved a record low 6~u.c. dead-layer by removing a sizable part of the oxygen vacancies. Our results give a comprehensive understanding of the dead-layer behavior in LSMO/STO, and shed light on other interface effects in the strongly correlated material. Furthermore, it provides the possibility that, with the capability to engineer the oxide films at atomic scale, we could investigate the strongly correlated system more deeply and ultimately handle the double-edged complexity.

\textbf{Methods:} LSMO films were fabricated by atomic layer-by-layer growth mode on 5$\times$5$\times$0.5mm$^3$ STO substrate. During growth, La and Sr were co-deposited to get the uniform doping. Thin films were grown at $740\,^{\circ}\mathrm{C}$ measured by pyrometer. The base pressure of OMBE chamber was $2\times 10^{-10} mbar$ , and LSMO films were grown under $1\times 10^{-6} mbar$ pressure of distilled ozone. These growth conditions were optimized for the nominal LSMO to reach the highest Curie temperature. The resistivity was determined in a Van der Pauw four-probe configuration. The magnetic susceptibility measurements were conducted in a MPMS system made by Quantum Design. Photoemission data were taken \textit{in situ} with a SPECS UVLS discharge lamp (21.2eV He-I$\alpha$ light) and a Scienta R4000 electron analyzer, at 180~K for 6~u.c. insulating films to avoid charging effect, and 25~K for all others. \textit{In situ} LEED were taken with 100~eV incident beam energy.

\textbf{Acknowledgement:} The authors thank X. G. Gong, B. A. Davison,  Jak Chakhalian and D.W. Shen for fruitful discussion. This work is supported in part by the National Science Foundation of China, and National Basic Research Program of China (973 Program) under the grant Nos. 2012CB921400, 2011CB921802 and 2011CBA00112.

\textbf{Author contributions:} R.P., H.C.X., M.X., B.P.X. performed the in-situ photoemission measurement and LEED study. R.P. and H.C.X. designed the films. R.P., H.C.X., M.X., X.X. and D.F.X. fabricated the films, and conducted the transport measurement. R.P., H.C.X. and D.L.F. established the dead-layer model. H.C.X., R.P. and J.F.Z. set up the OMBE system. D.L.F. and R.P. wrote the paper. D.L.F. is responsible for the infrastructure, project direction and planning.

\textbf{Additional Information:} The authors declare no competing
financial interests. Correspondence and requests for materials
should be addressed to D.L.F. (dlfeng@fudan.edu.cn).

\end{document}